\documentclass[aps,preprint,showpacs,nofootinbib,12pt]{revtex4}
\usepackage{graphicx}
\usepackage{epsfig}
\usepackage[dvips]{color}

\begin{document}

\def\beq{\begin{eqnarray}}
\def\eeq{\end{eqnarray}}
\def\non{\nonumber}

\title{ Whether new data on $D_s\rightarrow f_0(980) e^+ \nu_e$ can be understood if
 $f_0(980)$ consists of only the conventional $q\bar{q}$ structure}

\author{ Hong-Wei Ke$^{1}$   \footnote{khw020056@hotmail.com},
         Xue-Qian Li $^{2}$  \footnote{lixq@nankai.edu.cn} and
         Zheng-Tao Wei$^{2}$ \footnote{weizt@nankai.edu.cn} }

\affiliation{
  $^{1}$ School of Science, Tianjin University, Tianjin 300072, China \\
  $^{2}$ School of Physics, Nankai University, Tianjin 300071, China }

\begin{abstract}
\noindent Only two isospin-singlet scalar mesons $f_0(600)$
($\sigma$) and $f_0(980)$  exist below 1 GeV, so that it is natural
to suppose that they are two energy eigenstates which are mixtures
of ${1\over\sqrt 2}(u\bar u+d\bar d)$ and $s\bar s$. Is this picture
right? Generally, it is considered that $f_0(600)$ mainly consists
of ${1\over\sqrt 2}(u\bar u+d\bar d)$, if so, the dominant component
of $f_0(980)$ should be $s\bar s$. The recent measurement of the
CLEO collaboration on the branching ratio of $D_s\rightarrow
f_0(980) e^+ \nu_e$ provides an excellent opportunity to testify the
structure of $f_0(980)$, namely whether the data can be understood
as long as it consists of mainly the conventional $q\bar q$
structure. We calculate the form factors of $D_s\rightarrow
f_0(980)$ in the light-front quark model (LFQM) and the
corresponding branching ratio of the semileptonic decay. By fitting
the data, we obtain the mixing angle $\phi$. The obtained mixing
angle shows that the $s\bar s$ component in $f_0(980)$ may not be
dominant.
\end{abstract}
 \maketitle

\section{introduction}

Since the SU(3) quark model of the hadrons was founded, dispute
about quark structure of resonances never ceases.  Recently, it
turns out to be a hot topic because many new resonances have been
experimentally observed, such as $X(3872)$, $X(3940)$, $Y(3930)$,
$Z(3930)$, $Y(4430)$, $Y(4140)$ and it seems hard to accommodate
them in the conventional quark-structures, i.e. meson consists of a
quark and an antiquark; baryon consists of three valence quarks
\cite{Li:2008ey}. The exotic features of the newly discovered mesons
suggest that they may be multi-quark states (tetraquark or molecular
states), hybrids and glueballs, especially may be mixtures of all
those possible states with the regular $q\bar q$ components. In
fact, the story began a long time ago, when Weinstein and Isgur et
al. suggested that $f_0(980)$ and $a_0(980)$ might be $K\bar K$
molecular states and dissolve into $K\bar K$ final states near the
kinetic threshold.

The mass and width of $f_0(980)$ are measured as $980\pm 10$MeV and
$40-100$ MeV \cite{PDG08} respectively, but its structure is still
obscure so far. $f_0(980)$  was identified as a four-quark state in
Ref. \cite{Jaffe} where the authors evaluated its mass by
postulating a $qq\bar{q}\bar{q}$ structure in terms of the MIT bag
model. In Ref. \cite{Jaffe2000} the authors investigated a
possibility that the light scalar meson $f_0(980)$ was a $\bar
q^2q^2$ state rather than $\bar{q}q$ based on a lattice calculation.
Instead, since the resonance is close to the $K\bar{K}$ threshold a
$K\bar{K}$ molecular structure was suggested by Weinstein and Isgur
\cite{KK}. In Ref. \cite{Mixing} a possibility that it is a mixture
of $q\bar{q}$, the so-called scalaron coupled to $K\bar{K}$, was
discussed. Moreover, it is also counted as a glueball
\cite{Glueball}. By analyzing the experimental measurements of the
concerned decay and production processes, some authors affirmed that
the conventional $q\bar{q}$ structure might tolerate the available
data \cite{Morgan:1974,chang:1989,kekez,phi,Anisovich,Bennich}.
Those diverse interpretations should eventually be negated or
confirmed by more accurate experimental measurements, as well as
further theoretical investigations.

From the theoretical aspect, only $f_0(600) \;(\sigma)$ and
$f_0(980)$ are isospin-singlet scalar mesons below 1 GeV, so it is
natural to suppose that they are partner energy eigenstates which
are on-shell physical particles and mixtures of components of a
complete basis. The simplest  basis for the iso-singlet quark
structure consists of two components: ${1\over\sqrt 2}(u\bar u+d\bar
d)$ and $s\bar s$. Thus, one may write \cite{kekez,phi,HYCheng}
\begin{eqnarray}\label{mixing}
 \left ( \begin{array}{ccc} f_0 (600)(?)\\  f_0(980) \end{array} \right )=
 \left ( \begin{array}{ccc}
  \rm \cos\phi & \rm -\sin\phi \\
  \rm \sin\phi & \rm \cos\phi \end{array} \right )
 \left ( \begin{array}{ccc} f_{0q} \\  f_{0s} \end{array}\right ),
 \end{eqnarray}
where $f_{0q}={1\over\sqrt 2}(u\bar u+d\bar d)$ and $f_{0s}=s\bar
s$. Here the question mark denotes that we cannot determine if the
partner of $f_0(980)$ is indeed $f_0(600)$ (see below for more
discussions). If we could obtain a general Hamiltonian matrix which
not only gives the diagonal elements, i.e.
$<f_{0q}(f_{0s})|H|f_{0q}(f_{0s})>$, but also the off-diagonal
elements $<f_{0s}(f_{0q})|H|f_{0q}(f_{0s})>$, then it would be easy
to diagonalize the matrix to obtain the physical eigenenergies and
mixing angle. However, unfortunately, since the matrix elements are
fully dominated by the non-perturbative QCD and cannot be derived
from our present knowledge on QCD yet, one needs to determine the
mixing angle by fitting the available data. The mixing angle was
obtained as $\phi\sim 18.3^\circ$ in terms of the process
$f_0(980)\to \pi\pi$ \cite{kekez} and the authors of Ref. \cite{phi}
got $\phi\sim 16^\circ$, then they renewed their value to $(23\pm
3)^\circ$ \cite{phi1}. It seems that these results indicate  $s\bar
s$ should be the dominant component of $f_0(980)$. In Ref.
\cite{Anisovich}, the authors analyzed two processes:
$\phi(1202)\rightarrow f_0(980)\gamma$ and $f_0(980)\rightarrow
\gamma\gamma$ and gained $\phi=(142\pm6)^\circ$.  By studying
non-leptonic decay of $D(D_s)\rightarrow f_0(980)$, El-Bennich $et\,
al.$ \cite{Bennich} obtained $\phi=(32\pm4.8)^\circ$  in terms of
the covariant light-front quark model (CLFD) \cite{CLFD} and
$\phi=(41.3\pm5.5)^\circ$ by the dispersion relation (DR) approaches
\cite{DR} respectively. Cheng $et\, al.$ \cite{HYCheng} found that
the mixing angle lies in the ranges of ($25^\circ<\phi<40^\circ $ or
$140^\circ<\phi<165^\circ$) in a phenomenological study.  The
numerical values on the mixing angle are so disperse, can we
conclude that $f_0(980)$ is not a pure $q\bar q$ bound state? It
demands an answer which should be coming from new measurements and
theoretical efforts.

Indeed, all of our information on the inner structure must be
obtained from experimental measurements. With great improvements in
experimental facility and detection technique, much more accurate
data on $f_0(980)$ have been achieved which help theorists to make
judgement if the available models are correct. The recent
measurement on semileptonic decay $D_s^+\rightarrow f_0 e^+ \nu_e$
by the CLEO collaboration may provide a unique opportunity to
testify the validity of the $q\bar q$ structure of $f_0(980)$
\cite{CLEO} \footnote{Here $f_0$ refers to $f_0(980)$
\cite{WeiKeCLEO}.}. The decay rate was measured as $(0.13\pm 0.04\pm
0.01)\%$.

The present work is to testify if the mixture ansatz can tolerate
the data of the semileptonic decay. Assuming the $q\bar q$ structure
of $f_0(980)$ as $\sin\phi\,{1\over \sqrt 2}(u\bar u+d\bar
d)+\cos\phi \,s\bar s$, we calculate the decay width. From the
experimental data, the mixing angle $\phi$ would be obtained. Then
we discuss whether it is consistent with the results obtained by
fitting other experiments.

It is easy to see from the quark diagram, that at the tree-level,
only the $s\bar s$ component contributes to the transition. The
amplitude at the quark level can be obtained in terms of the weak
effective theory, so the key point is how to more accurately
calculate the hadronic transition matrix elements. Here we employ
the light-front quark model (LFQM). LFQM is a relativistic model
which has obvious advantages when light hadrons are involved
\cite{Jaus:1999zv,Cheng:2003sm}. The light-front wave function is
manifestly Lorentz invariant and expressed in terms of the
light-front momentum fractions and the relative transverse momenta
which are independent of the total hadron momentum. Applications of
this approach can be found in Ref. \cite{Hwang:2006cua} and the
references therein.  The LFQM has been used to calculate the decay
rates of $D_s$ into $\eta$ and $\eta'$ which contain similar
structure $C_1(u\bar{u}+b\bar{b})+C_2 s\bar{s}$ \cite{Wei:2009nc}.
Now we extend our scope to study $D_s\to f_0(980)$. Even though
$\eta$ and $\eta'$ are pseudoscalar mesons and $f_0(600)$ and
$f_0(980)$ are scalars, their isospin structures are the same.

Concretely, we calculate the form factor of $D_s\rightarrow
f_0(980)$ and obtain the decay rate of $D_s\rightarrow f_0(980) e^+
\nu_e$ where the parameter $\phi$ is a free one. Comparing our
result with the experimental data we fix the value of the mixing
parameter $\phi$. With this mixing parameter, we can further predict
the branching ratio of $D^+\rightarrow f_0(980) e^+ \nu_e$ which
will be measured in the future on BES III. It is noted that if the
$q\bar q$ structure: $|f_0(980)>=\sin\phi |{1\over\sqrt 2}(u\bar
u+d\bar d)>+\cos\phi |s\bar s>$ is right, the above reaction can
only realize via the Cabibbo-suppressed diagram to which only the
$u\bar u+d\bar d$ component contributes, and/or the annihilation
diagram which is even further suppressed, so the ratio should be
smaller by one or two orders. The smallness of the branching ratio
should be a check for the structure of $f_0(980)$.

This paper is organized as follows: after the introduction, in
section II we will present the form factors of $D_s\to f_0(980)$
which are evaluated in LFQM, then we obtain $\phi$ by fitting the
experimental data, at the end of the section, we make a prediction
on the branching ratio of $D^+\rightarrow f_0(980) e^+ \nu_e$.
Meanwhile, we briefly estimate the errors which originate from both
experimental and theoretical aspects. Section III is devoted to our
conclusion and  discussions.

\section{Study on the process involving $f_0$ mesons in  LFQM}
\begin{figure}
\begin{center}
\begin{tabular}{cc}
\includegraphics[width=9cm]{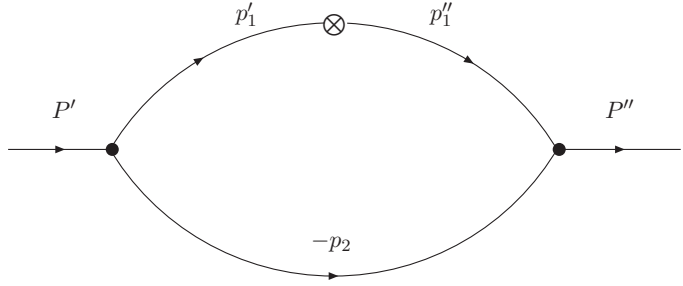}
\end{tabular}
\end{center}
\caption{ Feynman diagram for meson transition amplitude}
\label{fig:LFQM}
\end{figure}

Now let us calculate the form factors of  $D^+_s\rightarrow
f_0(980)$ in LFQM. Here we assume that $f_0(980)$ is of the
conventional $q\bar q$ structure $|f_0(980)>=\sin\phi |{1\over\sqrt
2}(u\bar u+d\bar d)>+\cos\phi |s\bar s>$.  The transition diagram is
given in Fig. \ref{fig:LFQM}.

\subsection{Formulations}\label{formula}
The form factors for $P\rightarrow S$ transition are defined as
\begin{eqnarray}\label{2s1}
\langle
S(P'')|A_\mu|P(P')\rangle=i\left[u_+(q^2)\mathcal{P}_\mu+u_-(q^2)q_\mu\right],
\end{eqnarray}
It is convenient to redefine them as
\begin{eqnarray}\label{2s2}
\langle
S(P'')|A_\mu|P(P')\rangle=-i\left[\left(\mathcal{P}_\mu-\frac{M'^2-M''^2}{q^2}q_\mu\right)
F_1(q^2)+\frac{M'^2-M''^2}{q^2}q_\mu F_0(q^2)\right],
\end{eqnarray}
where $q=P'-P''$ and $\mathcal{P}=P'+P''$. The relations between
them are
\begin{eqnarray}
F_1(q^2)=-u_+(q^2), F_0(q^2)=-u_+(q^2)-\frac{q^2}{q\cdot{
\mathcal{P}}}u_-(q^2).
\end{eqnarray}

Functions $u_{\pm}(q^2)$ can be calculated in LFQM and their
explicit expressions are presented as \cite{Cheng:2003sm}
\begin{eqnarray}\label{2s4}
u_+(q^2)=&&\frac{N_c}{16\pi^3}\int dx_2d^2p'_\perp
 \frac{h'_ph''_s}{x_2\hat{N}'_1\hat{N}''_1}
 \left[-x_1(M'^2_0+M''^2_0)-x_2q^2+x_2(m_1'+m_1''^2)\right.\nonumber\\&&
 \left.+x_1(m'_1-m_2)^2+x_1(m''_1+m_2)^2\right],\nonumber\\
u_-(q^2)=&&\frac{N_c}{16\pi^3}\int dx_2d^2p'_\perp
\frac{2h'_ph''_s}{x_2\hat{N}'_1\hat{N}''_1}
 \left\{x_1x_2M'^2+p'_\perp+m_1'm_2+(m_1''+m_2)(x_2m_1'+x_1m_2)\right.\nonumber\\&&
 -2\frac{q\cdot \mathcal{P}}{q^2}\left(p'^2_\perp+2\frac{(p'_\perp\cdot
 q_\perp)^2}{q^2}\right) -2\frac{(p'_\perp\cdot
 q_\perp)^2}{q^2}+\frac{(p'_\perp\cdot
 q_\perp)}{q^2}\left[M''^2-x_2(q^2+q\cdot P)\right.\nonumber\\
 &&-(x_2-x_1)M'^2 \left.\left.
 +2x_1M'^2_0-2(m'_1-m_2)(m'_1-m''_1)\right]\right\},
\end{eqnarray}
where $m'_1,m''_1$ and $m_2$ are the corresponding quark masses,
$M'$ and $M''$ are the masses of the initial and final mesons
respectively. The wave function is usually chosen to be Gaussian and
the parameter $\beta$ in the Gaussian wave function determines the
confinement scale and is expected to be of order $\Lambda_{\rm
QCD}$.  All other notations are given in the appendix. Some
parameters, such as $m_s=0.37$ GeV, $m_c=1.4$ GeV are taken from
Ref. \cite{Cheng:2003sm}; $\beta_{D_s}=0.592$ GeV and
$\beta_{D}=0.499$ GeV are fixed by fitting concerned processes
\cite{Wei:2009nc}.

In terms of Eq.(\ref{mixing}), the quark structure of $f_0(980)$
may be written as $|f_0(980)>=\sin\phi |{1\over\sqrt 2}(u\bar
u+d\bar d)>+\cos\phi |s\bar s>$. Since strange quark $s$ in $D_s$
directly transits into the final scalar meson, one can notice that
only $s\bar{s}$ component of $f_0(980)$ contributes to the
transition $D_s \to f_0(980)\bar l\nu_l$.

In order to calculate the relevant form factors we need to know
$\beta^s_{f_0}$. We cannot obtain $\beta^s_{f_0}$ directly from its
decay constant as we did for the pseudoscalars, because the decay
constant of $f_0$ is zero \footnote{The decay constant of $0^+$
state in LFQM can be written as
 \begin{eqnarray}
 f_s=\frac{N_c}{16\pi^3}\int
dx_2d^2p'_\perp\frac{h_s'}{x_1x_2(M'^2-M''^2_0)}4(m'_1x_2-m_2x_1).
 \end{eqnarray}
For $f_0(980)$, $m_1'=m_2$. The function $h_s'$ and other quantities
$M'^2$, $M''^2_0$ are symmetric functions of $x_1$ and $x_2$. Thus
the integration is zero, i.e. $f_{f_0}=0$.}. Following Ref.
\cite{Cheng:2003sm}, we set $\beta^s_{f_0}=0.3$ in our numerical
computations.

In the covariant light-front quark model, the calculation of form
factors is performed in the frame $q^+=0$ with $q^2=-q^2_{\perp}\leq
0$. Thus only the values of the form factors in the space-like
region can be obtained. The advantage of this choice is that the
so-called Z-graph contribution arising from the non-valence quarks
vanishes. In order to obtain the physical form factors, an
analytical extension from the space-like region to the time-like
region is required. The form factors can be parameterized in a
three-parameter form as
 \begin{eqnarray}\label{s14}
 F(q^2)=\frac{F(0)}{1-a\left(\frac{q^2}{M^2}\right)
  +b\left(\frac{q^2}{M^2}\right)^2}.
 \end{eqnarray}
where $F(q^2)$ represents the form factors $F_1,~ F_0$, and $F(0)$
is the form factors at $q^2=0$; $M$ is the mass of the initial
meson. The three parameters $F(0)$ and $a,~b$ are fixed by
performing a three-parameter fit to the form factors  which are
calculated in the space-like region and then extended to the
physical time-like region.

For semileptonic decay of a pseudoscalar meson($D$ or $D_s$) into
a scalar meson, i.e. $P(P')\to S(P'')l\nu$, the differential width
is
 \begin{eqnarray}
 \frac{d\Gamma}{dq^2}(P\to Sl\nu)=\frac{G_F^2|V_{CKM}|^2~p^3}{24\pi^3}|F_1(q^2)|^2,
 \end{eqnarray}
which is the same as the one for $P(P')\to P(P'')l\nu$
\cite{Zweber:2007zz}, where $q=P'-P''$ is the momentum transfer and
$q^2$ is the invariant mass of the lepton-neutrino pair; $p$ is the
final meson momentum in the $D$ or $D_s$ rest frame and
 \begin{eqnarray}
 p=|\vec{P''}|=\frac{\sqrt{\left(M^2-(M_f-\sqrt{q^2})^2\right)
  \left(M^2-(M_f+\sqrt{q^2})^2\right)}}{2M},
\end{eqnarray}
where $M_f$ denotes the mass of the produced meson. It is noted that
the differential  width is governed by only one form factor
$F_1(q^2)$ because we neglect the light lepton masses.

\subsection{Mixing angle in $f_0(980)$ and prediction on the decay rate of
 $D^+\to f_0(980)l^+\nu$}

We have calculated the form factor ${F_1}_{[{D_sf_0(980)}]}(q^2)$
and obtain ${F_1}_{[{D_sf_0(980)}]}(0)=0.434\cos\phi$ and
$a=1.03$, $b=0.267$. Obviously the mixing angle $\phi$ is included
in ${F_1}_{[{D_sf_0(980)}]}(0)$.

\begin{figure}
\begin{center}
\begin{tabular}{cc}
\includegraphics[scale=0.7]{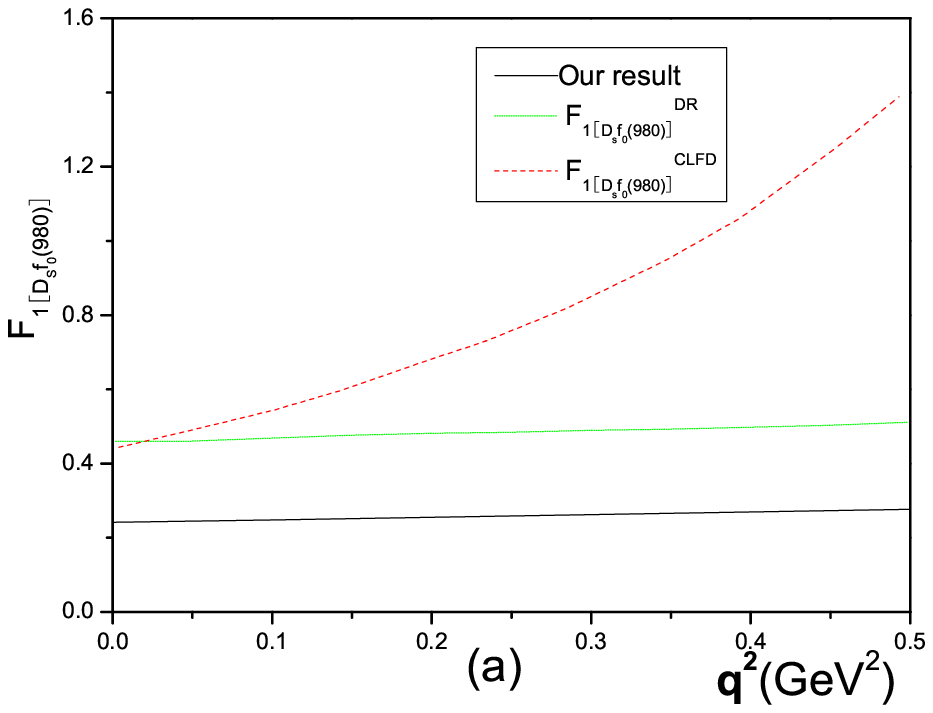}
\includegraphics[scale=0.7]{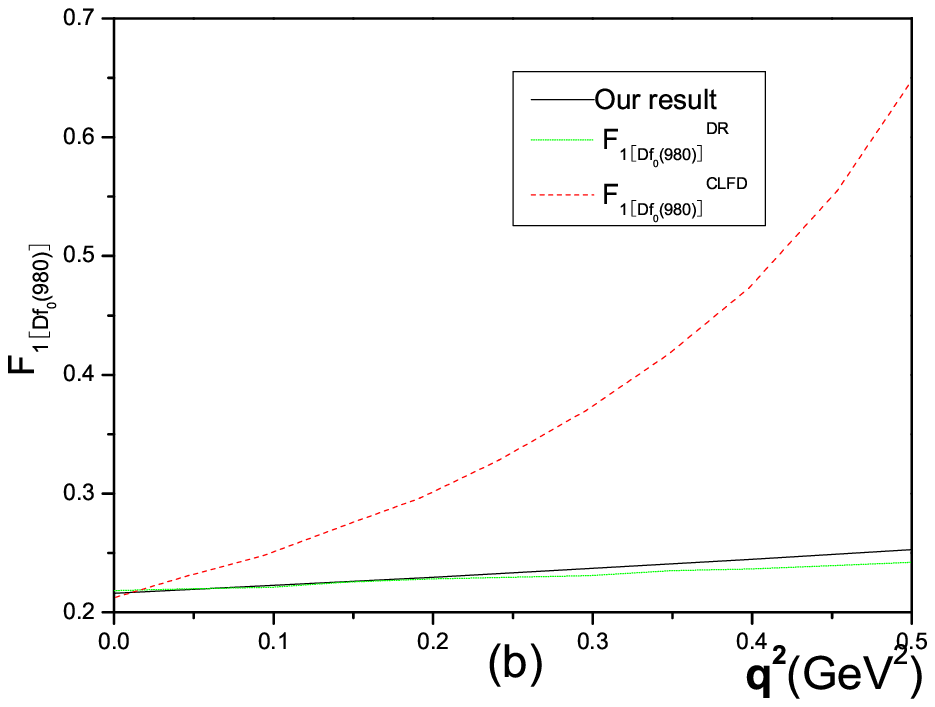}
\end{tabular}
\end{center}
\caption{Form factors of $|{F_1}_{D_sf_0(980)}(q^2)|$ and
$|{F_1}_{Df_0(980)}(q^2)|$ at different $q^2$. The solid lines
represent our results, dotted (in terms of CLFD) and dash-dotted (in
terms of the dispersion relation) lines are taken from
\cite{Bennich}. } \label{fig:fqs}
\end{figure}

With ${F_1}_{[{D_sf_0(980)}]}(q^2)$, we can compute the differential
width and branching ratio of $D_s^+\rightarrow f_0(980) e^+ \nu_e$
\begin{eqnarray}
 \mathcal{BR}(D_s^+\rightarrow f_0(980) e^+ \nu_e)=4.22\times 10^{-3}
  {\rm cos}^2\phi.
\end{eqnarray}
Comparing the theoretical evaluation of the branching ratio with the
experimental data we eventually obtain
 \begin{eqnarray}
 \phi=(56\pm 6)^\circ,
 \end{eqnarray}
and its symmetric angle in the second quadrant is
 \beq
 \phi=(124\pm 6)^\circ.
 \eeq
In the below discussions, we adopt the mixing angle in the first
quadrant for illustration, i.e. $\phi=(56\pm 6)^\circ$. This result
implies that $s\bar{s}$ is not the dominant component of $f_0(980)$
at all. The uncertainties in the numerical results originate from
the errors of the experimental measurements. Definitely, this result
contradicts to those given in \cite{kekez,phi} by at least two
standard deviations.

In fact, one needs the total width of $f_0(980)$ which unfortunately
was not precisely measured and spans a rather wide range from 40 to
100 MeV, to determine the mixing angle. In \cite{kekez}, the authors
used the lower bound of the width of $f_0(980)$ to fix the mixing
angle, so that there could be some flexibility in determining the
value. Even through taking into account the flexibility, there is
still a serious discrepancy.

We plot the form factors ${F_1}_{[D_sf_0(980)]}$ and
${F_1}_{[Df_0(980)]}$ in Fig.\ref{fig:fqs} where we make a
comparison of our results with that obtained in \cite{Bennich}.
Here ${F_1}_{[D_sf_0(980)]}^{CLFD}$,
${F_1}_{[D_sf_0(980)]}^{DR}$), ${F_1}_{[Df_0(980)]}^{CLFD}$ and
${F_1}_{[Df_0(980)]}^{DR}$) correspond to the form factors in the
processes $D_s\to f_0$ and $D\to f_0$ and calculated by the
authors of \cite{Bennich}, in terms of the covariant light-front
dynamics and dispersion relation approaches respectively, and the
subscripts explicitly mark the differences. We can see the shapes
of our results have the same $q^2$ dependence as $F_1^{DR}$,
moreover, our numerical result on the form factor
${F_1}_{[Df_0(980)]}$ is very close to ${F_1}_{[Df_0(980)]}^{DR}$,
whereas our ${F_1}_{[D_sf_0(980)]}$ is almost a half of
${F_1}_{[Df_0(980)]}^{DR}$.

Now let us  discuss the obvious difference between our results on
the form factor for $D_s\to f_0(980)$ and that given in
\cite{Bennich}. The difference comes from the different values of
$F(0)$. In Ref.\cite{Bennich}, the authors used the data of
non-leptonic decays of $D_s\to f_0(980)$ as input, instead, we
employ the data of semi-leptonic decays. A simple calculation may
support our results. In the figure, one notes that our
${F_1}_{[Df_0(980)]}$ is very close to ${F_1}_{[Df_0(980)]}^{DR}$,
and generally one can write
 \beq
 &&{F_1}_{[Df_0(980)]}(0)=C\sin\phi, \non\\
 &&{F_1}^{DR}_{[Df_0(980)]}(0)=C'\sin\phi',
 \eeq
where the unprimed quantities are ours and the primed ones are that
given in \cite{Bennich}. Thus we obtain
 \beq
 \frac{C}{C'}=\frac{\sin\phi'}{\sin\phi}\approx 0.79.
 \eeq
Similarly, expressions for $D_s\to f_0(980)$ are
 \beq
 &&{F_1}_{[D_sf_0(980)]}(0)=D\cos\phi, \non\\
 &&{F_1}^{DR}_{[D_sf_0(980)]}(0)=D'\cos\phi'.
 \eeq
So
 \beq
 \frac{\cos\phi}{\cos\phi'}\approx 0.74.
 \eeq
If one makes approximation $D/D'\approx C/C'$ which should be hold
in the flavor $SU(3)$ limit, he would immediately obtain
 \beq
 \frac{{F_1}_{[D_sf_0(980)]}(0)}{{F_1}^{DR}_{[D_sf_0(980)]}(0)}\approx 0.58.
 \eeq
This factor 0.58 can help to understand why our result of
${F_1}_{[D_sf_0(980)]}$ is only half of ${F_1}^{DR}_{[D_sf_0(980)]}$
shown in Fig.\ref{fig:fqs}(a). In fact, if we take the form factor
${F_1}^{DR}_{[D_sf_0(980)]}$ for calculating the branching ratio of
the semileptonic decay $D_s\to f_0(980)+l^++\nu$, the result would
be four times larger than the data.

Now, let us see what consequent implications to be obtained if we
use the obtained value. Based on the general knowledge of quantum
mechanics, there should exist another physical state $f_0'$ which is
orthogonal to $f_0(980)$.  There could be two choices, $i.e.$ the
supposed partner is lighter or heavier than $f_0(980)$. Even though
$\phi$ deviates from $\phi'$ for $\eta$ and $\eta'$ mixing angle
($39.3^\circ)$ \cite{Feldmann:1998}\footnote{ In literature, the
$\eta-\eta'$ mixing is defined as
 \begin{eqnarray}\label{as18}
 \left ( \begin{array}{ccc} \eta \\  \eta' \end{array} \right )=
 \left ( \begin{array}{ccc}
  \rm cos\phi' & \rm -sin\phi' \\
  \rm sin\phi' & \rm cos\phi' \end{array} \right )
 \left ( \begin{array}{ccc} \eta_{qq}\\  \eta_{ss} \end{array}\right )
 \end{eqnarray}
where $\eta_{qq}={1\over\sqrt 2}(u\bar u+d\bar d)$ and
$\eta_{ss}=s\bar s$.}, one may expect that the mass difference
between $f_0(980)$ and its partner should be somehow close to that
between $\eta$ and $\eta'$ i.e approximately 410 MeV, because in
both cases, the mass difference should be determined by the chiral
symmetry breaking \cite{etamixing}. If its partner is the lighter
than it, one is tempted to identify $f_0(600)$, however, the mixing
angle $\phi$ suggests that the partner of $f_0(980)$ possesses a
sizable $s\bar s$ component, and this allegation definitely
contradicts to our present knowledge on $f_0(600)$. Thus, one might
expect its partner to be a heavier one. According to this criterion
the resonance $f_0(1370)$ seems to be a candidate. From
\cite{PDG08}, we find that $f_0(1370)$ whose mass is measured from
1200 to 1500 MeV, indeed is plausible. We can estimate the branching
ratio of $D_s^+\rightarrow f_0(1370) e^+ \nu_e$ to obtain the angle
$\phi$ as
\begin{eqnarray}
 \mathcal{BR}(D_s^+\rightarrow f_0(1370) e^+ \nu_e)\propto {\rm
 sin}^2\phi.
\end{eqnarray}
The future experiment will measure the branching ratio of
$D_s^+\rightarrow f_0(1370) e^+ \nu_e$, and by fitting the new
data, we are able to determine the mixing angle $\phi$ again. If
the newly obtained value of $\phi$ is consistent with our value
$\phi=56.2^\circ$ within a certain error tolerance, it means that
the data support our postulate on the structure of $f_0(980)$ and
$f_0(1370)$, otherwise, we should turn to other possibilities.

If the flavor structure  of $f_0(980)$ is correct we can estimate
the decay rate of $D^+\rightarrow f_0(980)$ where only the $d\bar d$
component contributes and the transition is Cabibbo suppressed. For
that case, similar to the previous calculations, the three
parameters are achieved as ${F_1}_{[Df_0(980)]}(0)=0.216$ and
$a=1.16$, $b=0.25$ respectively. Following all the procedures we
made above, we achieve the branching ratio of $D^+\rightarrow
f_0(980) e^+ \nu_e$ as about $(5.7\pm 0.9)\times 10^{-5}$. Although
the branching ratio is very small as expected, we believe that the
future experimental facilities with high luminosity and precision
can do the job.

\subsection{Error analysis}

Now, we need to approximately estimate possible uncertainty in our
result. Even though the errors in the theoretical computations are
not fully under control, an approximately estimation would still be
possible and necessary.

Concretely, there are three error sources: the inherent uncertainty
in the theoretical model LFQM; the inputs to determine the model
parameters which are generally obtained by fitting several
well-measured processes; the data of CLEO which we are going to
employ to fix the mixing angle $\phi$.

Since as discussed above, the LFQM has been successfully applied to
analyze similar reactions, we have confidence that the higher order
effects are not important and the errors brought up by the model can
be neglected in comparison with that from other sources.

The second source is the errors in the inputs which cause
uncertainty in our theoretical calculation. To estimate the errors
we take a relatively easier way.   One notes \cite{Cheng:2003sm},
that the form factors $u_\pm$ are related to the form factors
$f_\pm$ in the $P\rightarrow P$ transition as,
\begin{eqnarray}
 u_\pm=-f_\pm(m''_1\rightarrow -m''_1, h''_p\rightarrow h''_s),
\end{eqnarray}
so it is natural to suppose that the theoretical uncertainties in
the LFQM calculations are the same for $P\rightarrow S$ and
$P\rightarrow P$. Obviously,  there may be a difference between
the form factors of $P\rightarrow S$ and $P\rightarrow P$ because
the wave functions of a scalar and a pseudo-scalar are not the
same (see Eq. \ref{app4} in appendix) $i.e.$ the  dependence of
those form factors on the parameter $\beta$ is different. But at
least they should have the same order of magnitude, thus we are
able to estimate the uncertainties in the form factors of
$P\rightarrow S$ by transferring that for $P\rightarrow P$.

We need to study the sensitivity of results to $\beta_{D_s}$ ($i.e.
\,\,\beta'$) and $\beta_{f_0(980)}$ ($i.e.\,\, \beta''$).  For the
analysis, we vary $\beta_{D_s}$ and $\beta_{f_0(980)}$ up and down
by 10\%,  then the theoretical uncertainties corresponding to
variation of $\beta_{D_s}$ and $\beta_{f_0(980)}$ are presented in
Table \ref{table1}. From the values, one can notice that with 10\%
up and down variations in $\beta$, the change of the mixing angle
falls within a range of 10\%. Thus, we can conclude that the results
are not very sensitive to $\beta$.

\begin{table}
\caption{Sensitivity to the variations in $\beta$ where $F_1(0),\ a
,\ b$ are defined in Eq. (\ref{s14}).} \label{table1}
\begin{ruledtabular}
 \begin{tabular}{ccccccccc}
 $\beta$ &$F_1(0)$ ( $10^{-3}$)&$a$& $b$&$\phi$&errors( $\phi-56.2^\circ$)\\\hline
 $\beta_{D_s}=0.652,~\beta_{f_0(980)}=0.30$ &-0.400cos$\phi$&0.947&0.208 &$52.5^\circ$&$-3.7^\circ$\\\hline
 $\beta_{D_s}=0.532,~\beta_{f_0(980)}=0.30$ &-0.468cos$\phi$&1.111&0.347 &$59.1^\circ$&$2.9^\circ$\\\hline
 $\beta_{D_s}=0.592,~\beta_{f_0(980)}=0.33$ &-0.479cos$\phi$&1.016&0.233 &$59.7^\circ$&$3.5^\circ$\\\hline
 $\beta_{D_s}=0.592,~\beta_{f_0(980)}=0.27$ &-0.382cos$\phi$&1.025&0.307 &$50.7^\circ$&$-5.5^\circ$\\
\end{tabular}
\end{ruledtabular}
\end{table}

For $D_s$ decays we employ the transition $D_s^+\to\eta e^+\nu_e$ to
estimate  errors in our theoretical calculation on the branching
ratio of $D_s^+\to f_0(980) e^+\nu_e$. Comparing our theoretically
calculated branching ratio of $D_s^+\to\eta e^+\nu_e$ in LFQM
$2.25\%$ \cite{Wei:2009nc} with the experimental data $(2.48\pm
0.29\pm 0.13)\%$ \cite{CLEO}, there is a $0.23\%$ deviation to the
central value.  Taking into account the experimental errors, the
integrated uncertainty of the theoretical estimation on the
branching ratio would be $\pm 0.32\%$ and the relative error is
$18\%$. We suppose the relative error for theoretical estimation on
the branching ratio of  $D_s^+\to f_0(980) e^+\nu_e$ to be the same
as that for $D_s^+\to\eta e^+\nu_e$, then by including the error of
the CLEO's data we obtain the whole uncertainty of $\phi$ as $\pm
7.2^\circ$.

Similarly, for   $D^+$ semi-leptonic decay, the decay of
$D^+\to\bar{K^0}e^+\nu_e$ is chosen to analyze the corresponding
error. The theoretical evaluation on the branching ratio is $9.74\%$
\cite{Wei:2009nc} while the experimental datum is $(8.6\pm 0.5)\%$
\cite{PDG08}, so the integrated uncertainty is $1.25\%$ in the
theoretical result and the relative error should be $12.8\%$. Then
as discussed above, the relative error of the branching ratio of
$D^+\rightarrow f_0(980) e^+ \nu_e$ estimated in LFQM should be $\pm
0.73\times 10^{-5}$.

Considering the error of the $\phi$ value the error in our
theoretical prediction for  semi-leptonic decay $D^+\rightarrow
f_0(980) e^+ \nu_e$ is estimated as  $\pm 1.3\times 10^{-5}$ and one
can expect $\mathcal{BR}(D^+\rightarrow f_0(980) e^+
\nu_e)=(5.7\pm1.3)\times 10^{-5}$.

\section{Discussions and conclusions}

Assuming that $f_0(980)$ possesses a regular $q\bar q$ structure,
namely, is written as $\sin\phi\,{1\over \sqrt 2}(u\bar u+d\bar
d)+\cos\phi \,s\bar s$, we calculate the branching ratio of the
semileptonic decay $D_s\rightarrow f_0(980) e^+ \nu_e$ in LFQM. To
fit the data which were measured by the CLEO collaboration, we
obtain the mixing angle $\phi$ as $\phi=(56\pm 7)^\circ$ or $(124\pm
7)^\circ$. This result implies that the $s\bar s$ component in
$f_0(980)$ is not the dominant one, instead, the fractions of
${1\over \sqrt 2}(u\bar u+d\bar d)$ and $s\bar s$ are almost equal.
It is definitely contradicts to the conclusion of \cite{kekez, phi}
 . Considering the experimental and theoretical errors, our
mixing angle is close to the results in \cite{Anisovich,Bennich}.

According to the basic principle of quantum mechanics, there should
exist its partner, i.e. another eigenstate of the Hamiltonian and
orthogonal to $f_0(980)$. A natural conjecture is that
$f_0(600)(\sigma)$ meson should be the most favorable candidate
because $f_0(600)$ and $f_0(980)$ are the only two isoscalar mesons
below 1 GeV. Obviously, it is against the previous studies where
$\sigma$ meson is confirmed to be mainly composed of $u\bar u+d\bar
d$, and the mixing angle with $s\bar s$ should be very small. Thus
if this conjecture is true, one should look for its partner above 1
GeV. As some authors suggested \cite{etamixing}, the mixing between
$\eta-\eta'$ is related to the chiral symmetry breaking and their
mass difference is determined by the values of the quark
condensates. The mass splitting is about 410 MeV. We suppose that
the mass difference of $f_0(980)$ and its partner should also be
related to the chiral symmetry breaking parameters, at least the
order of magnitude of the mass splitting should be close to that of
$\eta-\eta'$. As we take their mass difference as 410 MeV,
$f_0(1370)$ would be the favorable candidate. Surely, we cannot rule
out $f_0(1500)$ because its mass is not too far away at all. In this
work, we assume that $f_0(1370)$ is the isospin partner of
$f_0(980)$ and calculate the branching ratio of $D_s^+\rightarrow
f_0(1370) e^+ \nu_e$ which should be proportional to $\sin^2\phi$,
in the same framework. If the data which will be obtained by the
CLEO and/or BES collaborations, confirm this $\phi$ within a
reasonable error tolerance, our picture is supported, otherwise we
need re-consider the whole scenario.

Moreover, in the same theoretical framework, we estimate the
branching ratio of $D^+\to f_0(980)+l^++\nu$ and predict it as
$(5.7\pm 1.3)\times 10^{-5}$. Since in the process $D^+\to
f_0(980)+l^++\nu$ only the $d\bar d$ component contributes to the
transition (see the quark-diagram), the branching ratio is
proportional to $\sin^2\phi$, a measurement on it can help to
confirm the validity of the $q\bar q$ structure of $f_0(980)$.

As a conclusion, we obtain a mixing angle by fitting the new data of
CLEO which shows that $f_0(980)$ is not dominated by the $s\bar s$
component.

\section*{Acknowledgments}

This work is supported by the National Natural Science Foundation of
China (NNSFC) and the special grant for the PH.D program of the
Chinese Education Ministry. One of us (Ke) is also partly supported
by the special grant for new faculty from Tianjin University.

\appendix

\section{Notations}

Here we list some variables appeared in this paper.  The incoming
(outgoing) meson in Fig. \ref{fig:LFQM} has the momentum
${P'}^({''}^)={p_1'}^({''}^)+p_2$ where ${p_1'}^({''}^)$ and $p_2$
are the momenta of the off-shell quark and antiquark and
\begin{eqnarray}\label{app1}
&& p_1'^+=x_1P'^+, \qquad ~~~~~~p_2^+=x_2P'^+, \nonumber\\
&& p'_{1\perp}=x_1P'_{\perp}+p'_\perp, \qquad
 p_{2\perp}=x_2P'_{\perp}-p'_\perp,
 \end{eqnarray}
with  $x_i$ and $p'_\perp$ are internal variables and $x_1+x_2=1$.

The variables $M_0'$, $\tilde {M_0'}$, $h_p'$, $h_s'$ and
$\hat{N_1'}$ are defined as
\begin{eqnarray}\label{app2}
&&M_0'^2=\frac{p'^2_\perp+m'^2_1}{x_1}+\frac{p'^2_\perp+m^2_2}{x_2},\nonumber\\&&
\tilde {M_0'}=\sqrt{M_0'^2-(m_1'-m_2)^2}.
 \end{eqnarray}

\begin{eqnarray}\label{app3}
&&h_p'=(M'^2-M'^2_0)\sqrt{\frac{x_1x_2}{N_c}}\frac{1}{\sqrt{2}\tilde{M_0'}}\varphi',\nonumber\\&&
h_s'=(M'^2-M'^2_0)\sqrt{\frac{x_1x_2}{N_c}}\frac{\tilde{M_0'}}{2\sqrt{6}M_0'}\varphi'_p
 \end{eqnarray}

\begin{eqnarray}\label{app5}
\hat{N_1'}=x_1(M'^2-M'^2_0).
 \end{eqnarray}

where
\begin{eqnarray}\label{app4}
&&\varphi'=4(\frac{\pi}{\beta'^2})^{3/4}\sqrt{\frac{dp_z'}{dx_2}}{\rm
 exp}(-\frac{p'^2_z+p'^2_\perp}{2\beta'^2}),\nonumber\\&&
 \varphi'_p=\sqrt{\frac{2}{\beta'^2}}\varphi',
 \end{eqnarray}
with $p_z'=\frac{x_2M_0'}{2}-\frac{m_2^2+p'^2_\perp}{2x_2M_0'}$.

\end{document}